\documentstyle[11pt]{article}
\newcommand{\re}{\par\hangindent=0.5cm\hangafter=1\noindent}
 
 \newcommand{\etal}{{\it et al.~}}
 \newcommand{\alt}{\mbox{\raisebox{0.3ex}{$<$}\hspace{-1.1em}
                         \raisebox{-0.7ex}{$\sim$}} }
 \newcommand{\agt}{\mbox{\raisebox{0.3ex}{$>$}\hspace{-1.1em}
                         \raisebox{-0.7ex}{$\sim$}} }
 \newcommand{\Sec}[1]{\vspace{5mm}\begin{flushleft}{\bf #1 }
                     \end{flushleft}}
 \newcommand{\Subsec}[1]{\vspace{5mm}\begin{flushleft}{\it #1 }
                     \end{flushleft}}
 \newcommand{\Subsubsec}[2]{\vspace{5mm}\begin{flushleft}{#1 }{\it #2}
                     \end{flushleft} }
 
 \newcommand{\lcgs}{\rm erg~s^{-1}}
 \newcommand{\cc}{\rm cm^{-3}}
 \newcommand{\dif}[2]{{\partial #1 \over \partial #2}}
 
\newcommand{\ApJ}{{\it Astrophys. J.}}

\newcommand{\ApJS}{{\it Astrophys. J. Suppl.}}

\newcommand{\ARAAp}{{\it Annual Rev. Astron. Astrophys.}}
\newcommand{\PASJ}{{\it Publ. Astron. Soc. Japan}}

\begin{document}

{\Large
\begin{center}
{\bf Extended Hot Gas Halos Around Starburst Galaxies}\footnote{
submitted to \PASJ.}
\end{center}
}{\large
\vspace{5mm}
\begin{center}
K{\sc ohji} TOMISAKA\footnote{
e-mail address: tomisaka@ed.niigata-u.ac.jp}\\
Faculty of Education, Niigata University\\
8050 Ikarashi-2, Niigata 950-21\end{center}
\begin{center}
and\\
\end{center}
\begin{center}
J{\sc oel} N. BREGMAN\\
Astronomy Department, University of Michigan\\
Ann Arbor, MI\ 48109-1090, U.S.A.
\end{center}
\vspace{0.5cm}
\begin{center}
{\it Received 1992 August 12; accepted \hspace*{30mm}}
\end{center}
\vspace{1cm}
}
{\large
\begin{center}
{\bf Abstract}
\end{center}

Reanalysis of Einstein IPC data and new observations from the GINGA
 LAC indicate the presence of extended X-ray emission (10-50 kpc) around
 the starburst galaxy M82.
Here we model this emission by calculating numerical hydrodynamic
 simulations of the starburst event to much later times and larger scales
 than previously considered.
For our models, we adopt a supernova rate of 0.1 ${\rm yr}^{-1}$, and
 an extended low-density static halo that is bound to the galaxy.
There are three stages to the evolution of the wind-blown bubble and the
 propagation of the shock front: the bubble expands in an almost uniform
 density disk gas, with a deceleration of the shock front ($t \alt $ 3.6 Myr);
 breakout from the disk and the upward acceleration of the shock front (3.6
 Myr $\alt t \alt$ 18 Myr); propagation into the halo, leading to a more
 spherical system and shock deceleration (18 Myr $\alt t$).
For a halo density of $10^{-3} {\rm cm}^{-3}$, the outflow reaches a
 distance of 40-50 kpc from the center of the starburst galaxy in 50 Myr.
We calculate the time evolution of the X-ray luminosity and find that the
 extended starburst emits $3\times 10^{39}\lcgs$ to $10^{40}\lcgs$ in the
 GINGA LAC band and $\sim 10^{41}\lcgs$ in the Einstein or ROSAT
 HRI band.
The degree of the ionization equilibrium in the outflow and its effect
 on the iron K$\alpha$ line emission are discussed.

\vspace{3mm}

\newpage
\pagestyle{plain}

\Sec{1. Introduction}

Spiral galaxies with enhanced rates of star formation in their central
 regions have come under increasing scrutiny in an effort to understand
 the causes and evolution of this activity.
These systems, often referred to as starburst galaxies, possess
 considerable masses of molecular and atomic gas and
 are among the brightest sources of far infrared emission,
 presumably reradiated ultraviolet light from many early-type stars.
Based on the infrared luminosity, the CO mass, and the presence of
 many radio continuum point sources,
 it is estimated that a starburst galaxy such as M82 (NGC 3034) is
 producing massive stars that become supernovae at a rate of $\sim0.1$
 SN yr$^{-1}$ (Kronberg and Sramek 1985).
With this high rate of supernovae in a relatively small volume,
 the supernova remnants will merge quickly, creating a large bubble of
 hot gas in the center of the galaxy.

X-ray observations of starburst galaxies such as
 M82 and NGC253 reveal the presence of strong diffuse X-ray emission.
The Einstein Observatory HRI observations reveal that the emission
 extends perpendicular to the disk 2--3 kpc for
 M82 (Watson, Stanger, and Griffiths 1984; Kronberg, Bierman, and Schwab
1985)
 and 1--1.5 kpc for NGC 253 (Fabbiano and Trinchieri 1984).
The extended nature of this emission is indicative that the
 hot gas has been able to flow away from the central region of the galaxy.
Hydrodynamic models were developed to understand this outflow process
 in more detail and to reproduce the X-ray observations (Chevalier and
Clegg
 1985; Tomisaka and Ikeuchi 1988).
For various adopted initial gas distributions, Tomisaka and Ikeuchi (1988)
 calculated the evolution of the hot gas component,
 where in the early stages, the hot medium is confined to a bubble.
Eventually, this bubble elongates perpendicular to the disk
 (along the steepest density gradient) and begins to flow outward
 in a columnated fashion;
in these calculations, the hot gas is followed to a distance of 2--3 kpc
 from the midplane of the disk.
These results appear to be successful in that Einstein Observatory X-ray
 data can be understood within the context of the model.

Recent observations regarding the extent of the X-ray emission have
provided
 surprising results that indicate our understanding of the starburst
phenomenon
 is incomplete.
Einstein IPC observations (Fabbiano 1988) indicate that fainter X-ray
emission
 extends to about $\sim$ 10 kpc from the center of M82, but recent
observation
 with the Ginga X-ray satellite revealed the unexpected result that weak
 diffuse X-ray emission extends to distances of several tens of kiloparsecs
 (Tsuru \etal 1990).
It is this extremely extended X-ray emission, which was not anticipated in
the
 published models, that we investigate here with a series of gas-dynamical
 models that follow the starburst phenomenon to late times and include
such
 effects as an extended halo (or intergalactic) medium.
Although these calculations are aimed at understanding the recent Ginga
 observations, they are presented in a more general fashion so as to be
 applicable to a broader range of issues.

In the next section, the model of the galaxy and the initial density
 distribution are presented as well as the adopted numerical method.
In section 3, numerical results are described, while section 4 is devoted to
 the discussion of the expected X-rays emission, including the properties
 of the iron K$\alpha$ line ($h\nu\simeq$6.7 keV).

\Sec{2. The Model and the Numerical Method}

\Subsec{2.1. The Initial Density Distribution}

As in Tomisaka and Ikeuchi (1988; hereafter Paper I),
 we assume that the gravitational field is represented by
 a spherical stellar component with a modified King density distribution
$$
 \rho_*(\varpi)=\frac{\rho_c}{\displaystyle
[1+(\frac{\varpi}{r_c})^2]^{3/2}},
                                                                 \eqno(2.1)
$$
where $\varpi$ is the distance from the center of the galaxy.
The core radius $r_c$ and the stellar density at the center $\rho_c$
 are determined to fit the rotation law.
The gravitational potential is expressed as
$$
 \phi_*(\varpi)=-4\pi G \rho_c \left[
  \frac{{\displaystyle \ln
\left\{\frac{\varpi}{r_c}+[1+(\frac{\varpi}{r_c})^2]^{1/2}\right\}}}
      { \frac{\displaystyle \varpi}{\displaystyle r_c} }-1 \right]. \eqno(2.2)
$$
We assume that the gas distribution possesses cylindrical symmetry and use
 a cylindrical coordinate system ($z$, $r$, $\varphi$) to describe
 the hydrodynamics of the gas flow.
For the initial conditions, we assume that a part of the radial component
of
 gravity $e^2(\partial\phi_*/\partial r)$ ($e < 1$)
 is balanced by a centrifugal force $v_\varphi^2/r$, with the remainder of
 the gravitational force balanced by a pressure gradient.
If one were to simply assume that the gas rotates rigidly on cylinders,
 then the resulting hydrostatic isothermal density distribution would be
$$
\rho=\rho_0\exp\left[-\frac{\phi_*(z,r)-e^2\phi_*(0,r)}{c_s^2}\right].
\eqno(2.3)
$$
However, a difficulty with this distribution is that at large distances
 ($\varpi \gg r_c$) the gravity is weak, so the balance between the
 centrifugal force and the inward-directed density pressure gradient
produces
 a strong funnel around the $z$-axis (see Paper I) which may not be
physically
 realistic.
Consequently, we relax the constraint leading to this result by introducing
 the factor $\exp[-(z/z_{d})^2]$, thus allowing for
 slower rotation in the halo than the disk,
 an effect that is observed in M82 by Sofue \etal (1992).
Then, the resulting initial density distribution used here is
$$
\rho=\rho_0\exp\left\{-\frac{\phi_*(z,r)-e^2\exp\left[-(z/z_{d})^2\right]
\phi_*
(0,r)}{c_s^2}\right\}. \eqno(2.4)
$$

The various parameters that define the gravitational potential are chosen
 so as to reproduce known observations.
To fit the CO rotation curve of Nakai (1986), we use the parameters
$$
 r_c\simeq 350 {\rm pc}, \eqno(2.5)
$$
$$
 M_*=4\pi \rho_c r_c^3 \simeq 10^9 M_\odot
  \left(\frac{v_{\phi~max}}{60 {\rm km s^{-1}}}\right)^2
  \left(\frac{r_c}{350~{\rm pc}}\right) e^{-2}. \eqno(2.6)
$$
To include the presence of a halo or intergalactic gas,
 we assume the presence of a hydrostatic hot isothermal component of the
form
$$
\rho_h=\rho_{h0}\exp\left\{-\frac{\phi_*(z,r)-e^2\exp\left[-(z/z_{d})^2
\right]\phi_*(0,r)}{c_{sh}^2}\right\}, \eqno(2.7)
$$
where $c_{sh}$ and $\rho_{h0}$ are the sound speed and central gas
density of
 this hot component; the total density distribution is shown in figure 1.
On the $z$-axis, the cool (disk) component dominates below $\simeq$
 2 kpc, above which the hot (halo) component is distributed.
Here we take  $c_s=30{\rm km~s^{-1}}$,  $c_{sh}=300{\rm
km~s^{-1}}$,
 $\rho_0=20 {\rm cm}^{-3}$, $\rho_{h0}=2 \times 10^{-3} {\rm
cm}^{-3}$,
 $z_{d}=5$ kpc and $e=0.9$.
With this high value of $c_{sh}$ relative to the gravitational potential,
 $\rho_h$ is almost constant.
Due to the high halo temperature $\sim 6.5\times 10^6$K and the low density
 $\sim 2 \times 10^{-3} \cc$, the cooling time in the halo $\sim 2\times 10^9
 {\rm yr} (T/6.5\times 10^6{\rm K})(n/2 \times 10^{-3} \cc)^{-1}
 (\Lambda /2\times 10^{-23} {\rm erg~ cm^6~ s^{-1}})^{-1}$, is longer than the
  typical time scale of the computation $\alt 10^8$ yr.

\Subsec{2.2. Basic Equations and Numerical Method}

The two-dimensional, cylindrical symmetric, hydrodynamical equations are
the basic equations, which are given by
$$
\dif{\rho}{t}+\dif{}{z}(\rho v_z)+\frac{1}{r}\dif{}{r}(r\rho v_r)=0,
 \eqno(2.8)
$$
$$
\dif{\rho v_z}{t}+\dif{}{z}( \rho v_z v_z) +
 \frac{1}{r}\dif{}{r}(r\rho v_z v_r)= -\dif{p}{z} + \rho g_{z}^*,
 \eqno(2.9)
$$
$$
\dif{\rho v_r}{t}+\dif{}{z}( \rho v_r v_z) +
 \frac{1}{r}\dif{}{r}(r\rho v_r v_r)= -\dif{p}{r} + \rho g_{r}^*,
 \eqno(2.10)
$$
$$
  \dif{\epsilon}{t}+\dif{}{z}(\epsilon v_z)+\frac{1}{r}\dif{}{r}(r \epsilon
v_r)=-\Lambda(T) n^2-p (\dif{v_z}{z}+\frac{1}{r}\dif{r v_r}{r}),
                    \eqno(2.11)
$$
with
$$
\epsilon=\frac{p}{\gamma-1}, \eqno(2.12)
$$
where $\Lambda$ represents the radiative cooling coefficient
 (we use the cooling
 function derived by Raymond, Cox, and Smith 1976), ${\bf g}^*$ is the
 effective potential that includes the effect of the
 centrifugal force, and the other symbols have their usual meanings.

As the rotation of the gas was shown to be unimportant (Paper I), we do
not
 solve the equation of motion for the rotational component $v_\varphi$.
The effect of the rotational velocity is to reduce the gravity, leading to
 the effective gravity ${\bf g}^*=\nabla p/\rho$ given for the initial
 gas distribution.
The $r$-component of this effective gravity coincides
 with the ordinary effective potential as
$$
g_r^*=-\dif{\phi_*}{r}-\frac{v_\varphi^2}{r},  \eqno(2.13)
$$
 although it should be noted that ${\bf g}^*$ is no longer irrotational.

The ``monotonic scheme'' (van Leer 1977; Norman and Winkler 1986) was
adopted
 to solve the hydrodynamical equations, i.e., equations (2.8)--(2.12), where
 the advection and source terms are calculated separately.
For numerical stability at the shock front, a tensor artificial viscosity is
 included in equations (2.9)--(2.11) (see Tomisaka (1992)).
A cooling function is fitted to the Raymond-Cox-Smith function using a
third-order
 spline interpolation.
To include the radiative cooling, the source step of the energy equation (2.11)
 is solved by two steps, where the first step is
$$
  \frac{\tilde{\epsilon}-\epsilon^{(n)}}{\Delta t}
  =-{\rm work\ done\ by\ artificial\ viscosity},  \eqno(2.14)
$$
and the second step is
$$
  \frac{\epsilon^{(n+1)}-\tilde{\epsilon}}{\Delta t}=
  -\frac{1}{2}(p^{(n)}+p^{(n+1)}){\rm div}{\bf v}^{(n)}-
   n^{(n)~2} L(T^{(n+1)}),  \eqno(2.15)
$$
where the superscript refers to the time-step; the cooling function is
 rewritten as $\Lambda(T,~n)=n^2L(T)$.
Since the $(n+1)$ quantities enter the second step in a nonlinear fashion
 (implicit scheme),
 this equation must be solved iteratively to obtain the new thermal energy
 $\epsilon^{(n+1)}=p^{(n+1)}/(\gamma-1)=nkT^{(n+1)}/(\gamma-1)$.
Although we employ an implicit scheme to solve the energy equation,
 this scheme is totally vectorized and runs very efficiently on
 supercomputers.

\Sec{3. Results}

\vspace{-5mm}
\Subsec{3.1. Characteristics of Flow}

The numerical results of outflow are calculated using the conditions
 believed to prevail in the central region of M82,
 $r_{\rm SN}=0.1 {\rm yr}^{-1}$, and $n_0=20 {\rm cm}^{-3}$
 (other parameters are given in section 2).
The time evolution of the gas density, pressure, and velocity are shown
 in contour maps in figure 2, while the quantitative distribution of
 these quantities along the $z$-axis are presented in figure 3.
As the calculation progressed, a larger scale was required, so at early
 times (figure 2a) the cell size was
 $\Delta z= \Delta r=$ 25 pc and the number of cells was $(n_z,
n_r)=(500,250)$,
 at intermediate times (figures 2b and 2c) the cell size was
 $\Delta z= \Delta r=$ 50 pc and grid size was $(n_z, n_r)=(500,250)$,
while
 at later times, $\Delta z= \Delta r=$ 50 pc and the number
 of cells was 1000 by 500.

{}From the previous calculations on the outflow from starburst nuclei (Paper I)
 and superbubbles driven by supernova explosions (for review see
 Tenorio-Tagle and Bohdenheimer 1988 and Tomisaka 1991),
 a hot spherical bubble is initially produced at the center of the starburst
 but soon it begins to elongate
 in the $z$-direction when it becomes larger than 2--3 times the
 cold gas scale-height.
The effect is clearly seen in figure 2a and in Paper I
 (figures 4a, and 4b); the cold gas half thickness is related to $z_d$.
During this stage of the evolution, the usual double-shock structure is found
 (figure 3a) where matter ejected from stellar winds and supernovae pass
through
 an inward-facing shock (Weaver \etal 1977).
The swept-up material pushes into the surrounding region, driving an
 outward-going shock into the cool material.
This shock is radiative so the hot bubble is surrounded by a cooled shell
 of material.

The structure of the hot gas begins to change at an age of about $t=10$
Myr,
 when the upper part of the bubble encounters the low-density halo region
 ($n\sim 2 \times 10^{-3} {\rm cm}^{-3}$ and $nT \sim 10^4 {\rm
cm^{-3}~K}$),
 where the radiative cooling time is much longer than the
 expansion time of the bubble.
Furthermore, the shock begins to accelerate upward when it enters the low
 density material, and this leads to fragmentation of the cool shell, as is
 seen in figure 2b, where the open shell has a funnel shape.
Because of the long cooling time, a new cool shell at the top of the flow
 does not form, although there is still a cool shell in the lower part of the
 bubble that is in the disk.
At this stage, the upward flow is well-columnated and the gas moves
 almost parallel to the $z$-axis at supersonic speeds (figure 2b).
The flow is supersonic along the $z$-axis for $z \agt$ 1 kpc, it
 decelerates at the interface between the bubble and the undisturbed
medium,
 and then flows outward, which contributes to the tangential growth of
 the bubble.
The other contributor to the tangential expansion is the non-negligible
 azimuthal velocity component of the flow from the disk.

The essential condition that leads to a bipolar outflowing structure
 is the initial density stratification perpendicular to the galactic disk.
This was shown in Paper I by comparing the model with an initial density
 distribution similar to that adopted in the present paper to a model
 with an exact plane-stratified density distribution.
Although our initial conditions help promote a bipolar structure,
 the bipolar flow is nearly as pronounced in the calculations with
 a plane-stratified distribution.

 In the next phase of the starburst phenomenon (18 Myr, figures 2c, 3c), the
 structure and dynamics of the hot gas are dominated by the extended
 low-density medium.
A new feature, which is common in simulations of radio jets, is the
 appearance of shocks along the $z$-axis that repressurize and columnate
the
 flow.
At the base of the flow, from $(z,r)\simeq$(0kpc, 0kpc) to (5kpc, 0kpc),
 the flow
 experiences free supersonic expansion, but at a height of 5kpc, the
pressure
 of this flow region falls below that of its surroundings and a shock ensues,
 reheating the gas to $\agt 10^7$K.
Not only does a Mach disk appear, but a shock also occurs along the sides
 of the flow and may be traced upward from $(z,r)\simeq$(0kpc, 1kpc) to
the
 Mach disk near $z\simeq 5$ kpc.
 This shock is reflected at the symmetry $z$-axis via Mach reflection
 and the weak reflected shock can be traced further upward.
Following this first Mach disk, the flow expands once again, losing
 pressure and leading to another (weaker) repressurization shock
 along the $z$-axis.
This second shock can be traced from $(z,r)\simeq$(10kpc, 0kpc) to
 (5kpc, 3kpc), with a sharp change of direction near (10kpc, 2kpc).
At heights beyond 10kpc, the jet moves at constant velocity and
 with little expansion in the azimuthal direction.
This part of the flow ($z$ = 10-20kpc) is supersonic with respect to
 the hot halo.
As a result, since the gas flows upward supersonically, another
 outward-facing shock is formed in the ambient intergalactic matter
 and inside the shock front the ambient matter accumulates (see figure 3c;
 the distribution of mach number for $v_z$ clearly indicates
 several shocks and it shows that the flow at $z\agt $10 kpc is supersonic).
There is also a significant azimuthal component to part of the flow, which
 is caused by the flow through the oblique shocks that come off the Mach
disks.
This component of the flow is partly responsible for azimuthal expansion
of the
 hot bubble, making it more nearly spherical.

In the final stage of development, at 50 Myr (figures 2d and 3d), the
 disturbance triggered by supernova explosions
 reaches a height of $z\simeq$ 40 kpc from the starburst galaxy.
At this time, the hot gas would entirely envelope the disk of the galaxy.
As at the preceding time, two repressurizing shocks occur, although further
 out from the origin, and much of the flow is along the $z$-axis, but less
 narrowly focused.

These simulations possess three dynamical phases of the hot bubble that
 are associated with its evolution from the disk into the extended
 low-density halo.
By examining the height of the top of the outer-most shock front, $z_s$,
 against the
 time elapsed after the supernova burst began (figure 4), we see that the
 shock front decelerates during the first phase when the expansion occurs
 in the uniform disk material
 ($\partial^2z_s/\partial t^2 <0$ for $t \alt 3.6$Myr).
As the shock front pushes out of the disk and encounters a steep density
 gradient and low density material perpendicular to the disk
 (3.6Myr$\alt t \alt$18Myr), it accelerates and it is at this time
 that the initial dense shell fragments.
The final deceleration phase occurs because the bubble begins to expand
 in the halo that has a uniform density ($t\agt$18Myr).
The final expansion rate of this enormous hot bubble is similar to that
 of a pressure-driven superbubbles and is by given by $z_s\propto t^{3/5}$
 (Weaver \etal 1976).
If the flow were spherically symmetric, the size would be
$$
 R\simeq 31 {\rm kpc}
         \left(\frac{n_{\rm halo}}{2\times 10^{-2}\cc}\right)^{-1/5}
         \left(\frac{L}{3\times 10^{42}{\rm erg~s^{-1}}}\right)^{1/5}
      \left(\frac{t}{50 {\rm Myr}}\right)^{3/5}, \eqno(3.1)
$$
where $L$, and $n_{\rm halo}$ mean, respectively,
 the mechanical luminosity of supernova explosions $E_{\rm SN} r_{\rm SN}$,
 and the number density of the intergalactic medium.
However, the flow is actually non-spherical and is focused towards
 the $z$-axis.
The height of the actual outflow is larger than that expected assuming
 spherical symmetry ($R$).
Nevertheless, the geometric mean of the height of the top ($z_s$)
 and the half width ($r_{\rm max}$) of the bubble $(z_s r_{\rm max})^{1/2}$
 agrees well with the above estimate ($R$) and it shows that the expansion
 is proportional to $\propto t^{3/5}$.
Although the outflow shows the bipolarity due to the initial gas distribution,
 the dynamics of this process seems similar to the wind-blown bubble.
Previous calculations (e.g., Paper I) have studied the bubble development
as far as the acceleration phase, but the late stages of development have
not previously been investigated in the starburst context.

\Subsec{3.2. The Effect of Halo Pressure and Density}

To understand the effect of the physical state in the halo to the outflow,
 models with a range of different halo parameters were calculated.
Model B has a halo with a temperature and pressure that is ten times
 smaller than in model A but with the same density distribution.
The gravitational field is balanced by the pressure gradient.

Figure 5 shows the structure of this model at $t=$50 Myr.
Compared with figure 2d, we see that the overall structure
 is similar in the two models.
The expansion in the $z$-direction is almost identical in the two models,
 a result that is probably due to the importance of the upward-moving
 momentum and that the halo density distributions are the same.
However, the expansion in the $r$-direction shows differences in that the
 maximum radial size in model B is $\simeq$ 0.8 times smaller than in
 model A.
Comparing figure 5b (cross-cut view of figure 5a) with figure 2d,
 the pressure near the expanding front of this model is approximately
 a factor 30\% smaller than that of the previous model, which is a result
 of differences in the initial halo temperatures in the two models.
Thus, due to the decrease in the driving force, the expansion
 in the $r$-direction is slower than in the previous model.

To investigate this further, we calculated another model (C), in which
 the halo has the same density distribution of previous models but has
 a temperature and pressure that is 10 times higher than in model A.
Once again, the expansion in the $z$ direction is largley unaltered, with
 model C showing a slightly faster expansion than model A (10\%) at
 $t=$20My.
In contrast, the radial expansion is much faster than model A.
At $t=$20Myr, the maximum radial size of the outflow reaches $\sim$20 kpc,
 while in model A it is only $\sim$12 kpc.

In model D we have investigated the effect of changing the halo density
 by calculating a model with the same pressure as in model A but
 with one-tenth the density.

As expected, the outflow in a low-density halo expands more quickly
 than model A once the bubble has expanded into the halo ($t\agt$5Myr;
 figure 4).
While the expansion in the $z$-direction for models A and B are almost
 identical,
 the deceleration in the final phase becomes inefficient
 due to the lower density in model D.

{}From these models A--D, we concluded that the expansion in the $z$-direction
 is sensitive to the halo density but less sensitive to the halo pressure.
In contrast, the expansion in the $r$-direction is affected by both the
 halo pressure and density in the sense that the radial expansion becomes
 faster with higher halo pressure or with lower halo density.

\Sec{4. Discussion}
\vspace{-5mm}
\Subsec{4.1. Expected X-rays --- Comparison with Observation}

In this subsection,
 we calculate the expected X-ray emission by the hot gas outflowing
 from starburst galaxies.
We assume collisional equilibrium in the outflow and use the
 emissivity calculated by Raymond \etal (1976).
Assuming that the X-ray emission is optically thin, the
total luminosities of various X-ray bands are calculated by
$$
  L_{\rm Band}=2\int_{r=0}^{r=r_{\rm out}}\int_{z=0}^{z=z_{\rm out}}
   \Lambda_{\rm Band} n^2 2 \pi r dr dz,  \eqno(4.1)
$$
 where $\Lambda_{\rm Band}$ is a band X-ray emissivity coefficient
 while $z_{\rm out}$ and $r_{\rm out}$ are the sizes of the numerical
 outer boundaries.
In figure 6, the expected X-ray luminosity for model B is shown as a function
 of time for three energy band:  0.155--24.8 keV (solid line), 0.284--1.56 keV
 (dotted line), and 1.56--8.265 keV (dashed line).
It is assumed that the 0.284--1.56 keV band approximates the Einstein
 Observatory HRI band and that of 1.56--8.265 keV approximates the
 Ginga LAC band.
Fore- and background emission were subtracted from the value
 derived by equation(4.1), assuming that the expected X-ray luminosities
 at the initial state should vanish.
(Before the supernova explosions begin, the expected luminosities in the
 HRI and LAC bands from the galaxy and the extended halo are estimated
 to be $1.7\times 10^{39}\lcgs$ and $6.8\times 10^{35}\lcgs$, respectively.)

Before 5 Myr, the X-ray luminosities decrease, an effect that is due
 to the progressively decreasing density as the bubble expands and
 the shock front moves upward (this phase was studied in detail in
 Paper I).
In the subsequent $\simeq$5 Myr, the X-ray luminosities become
 a factor of 5 -- 10 brighter than the minimum value at $t\simeq$ 5 Myr.
This brightening occurs when the shock begins to accelerate ($t\simeq 3.6$Myr)
 which leads to a higher post-shock temperature and a higher luminosity.
After 10 Myr, the LAC band X-ray luminosity stays nearly constant,
 while in the lower energy bands, the luminosity increases slowly.
The fact that, at late times, the X-ray luminosities do not decrease
 monotonically but remain fairly constant is important observationally
 in that one expects to see this phenomenon over a long period of time,
 although at progressively lower surface brightnesses.

To more deeply understand the evolution of the X-ray luminosity, we first
 note that the X-ray emission is a combination of the shock-heated matter
 originally ejected from the parent galaxy and the halo material
 compressed by the outward-facing shock front.
If we adopt a simple spherically symmetric approximation for the evolution
 of the material, this hot gas drives a
 shock into the halo gas.
The time averaged temperature of the shock-heated material ejected from
 supernovae is $ T_{in} = (1/5) V_w^2 (\mu m_p/k)$, where $V_w$
 denotes the wind velocity and it is equal to
 $V_w=(2 r_{\rm SN}E_{\rm SN}/\dot{M}_{\rm SN})^{1/2}$ (Weaver
 \etal 1977).
This temperature, which remains constant during the course of the starburst
 event, is $T_{in}\simeq 10^8$K for parameters of our models A and B.
(This temperature seems to depend upon the geometry:
 in the actual starburst, a one-dimensional outflow like a funnel is formed
 and the temperature is below $10^7{\rm K}$,
 while in the spherical symmetric configuration the temperature must
 be $\sim 10^8$K.
This is because in an oblique shock the fraction of kinetic energy in the
 post-shock region is larger than that of the perpendicular shock.)
The shock speed decreases with time as it progresses further into the halo
 gas, so the post-shock temperature of the compressed halo matter
 decreases monotonically.
In addition to the evolution of the hot starburst event, the observed X-ray
 emissivity also depends on the temperature behavior of the bandpass
 cooling function (Raymond \etal 1976):  $\Lambda_{\rm HRI}$ is a
 decreasing function of $T$ for $T \agt 10^7$K and remains constant for
 $2\times 10^6 {\rm K} \alt T \alt 10^7 {\rm K}$, while $\Lambda_{\rm
 LAC}$ is an increasing function of $T$ for $T \alt 10^7$K.
Most of the emission of the HRI band seems come from the compressed
 halo matter.
Since the shock is decerelated and the post-shock temperature is decreasing
 from $\sim 10^7$K (after 10 Myr),
 the emission in the HRI band \underline{increases} due to the increase of
 $\Lambda_{\rm HRI}$.
On the other hand, since the emissivity decreases after  $\sim$ 10 Myr,
 the luminosity of the LAC band does not increase but remains constant,
 even if the amount of the accumulated mass increases.

Tsuru \etal (1990) reported that the luminosity emitted from the extended
 region surrounding M82 in the 2--10 keV band is approximately
 $\sim 4 \times 10^{39}\lcgs$, which corresponds to $\sim$ 10 \%
 of the total luminosity from M82.
In comparison, the X-ray luminosity calculated here, $3 \times 10^{39} -
 10^{40} \lcgs$, is consistent with the observed value.
Furthermore, the size of the X-ray emission region reaches
 $\simeq$ 50 kpc in half width and the size is expected to continue to grow
 without decreasing X-ray luminosity, provided that there is a halo or
 intergalactic medium with the same density at these larger sizes.
This extended X-ray emission is consistent with the observational width as
 seen by the LAC, which is wider than the point-spread function of LAC
 (FWHM of LAC is 1.1 degree and it corresponds to the linear size of
 $57(D/3.25{\rm Mpc})$ kpc, where $D$ represents the distance of the
 object).
Using our calculations, the mass of the hot gas is estimated to be
 $M_{\rm hot} \sim 1.5 \times 10^8 M_\odot (\tau/50 {\rm Myr})$.

The mass loss rate of model B corresponds to the assumption that
 the amount of ejected mass per supernova explosion $\Delta M$
 is equal to 30 $M_\odot$.
This value of the mass includes not only the ejecta of supernovae,
 but other sources, such as mass loss from stars that are not supernova
 progenitors or evaporated mass from clouds engulfed in the hot outflow.
We adopted this mass loss rate in models A--D.
However, we could have taken the limiting case that the only mass entering
 the system was the supernova ejecta, in which case the value of $\Delta
 M$ would be $\simeq 5 M_\odot$.
We have calculated a model (E) using this lower mass loss rate, in order
 to understand how the X-ray luminosity would be affected; the remaining
 parameters are identical to those in model B.
Since decreasing the mass loss rate leads to a higher input gas temperature,
 the energy flux into the system remains constant.
If a significant component of the X-ray emission were to come from this
 ejected matter, we might expect the X-ray spectrum to be harder.

The time variations of X-ray luminosities are shown in figure 7.
$L_{\rm LAC}$ decreases from $3\times 10^{39}\lcgs$ (model B) to
 $10^{39}\lcgs$ (model E), while
 $L_{\rm HRI}$ remains $\simeq 5\times 10^{40}\lcgs$, which seems to
 contradict the above expectation.
The X-ray luminosity in the HRI band must come largely from
 the accumulated halo/intergalactic matter, which depends on the total energy
 input to the system but is insensitive to the mass and temperature of the
 ejected matter.
If the X-ray emission in the LAC band is also from the accumulated
 intergalactic matter,
 the luminosity will be expected to be the same as that of model B.
Thus, a part of the luminosity in the LAC band is emitted from the ejected
 matter.
Since the ejected mass-loss rate is taken to be 1/6 that of
 model B, the emission measure of the ejected matter seems to become
 smaller than that of model B.
Even if we take into account that the emissivity $\Lambda_{\rm LAC}$
 becomes higher than that of model B, the total luminosity in the LAC
 band becomes smaller than from model B.

\Subsec{4.2. Iron Line Emission}

\Subsubsec{4.2.a. Iron Line Emission from the Starburst Galaxy}

Unfortunately, due to the low intensity of the extended emission,
 the spectrum of the extended emission around M82 was not detemined
 by the LAC on Ginga (Tsuru \etal 1990) and must be left to the next
 generation of X-ray telescopes, ASTRO-D, XMM or AXAF.
However, the spectra of the X-ray emission from the central part of
 nearby starburst galaxies, such as M82(Makishima and Ohashi 1989;
 Ohashi \etal 1990a) and NGC 253(Ohashi \etal 1990b), were obtained
 by Ginga.
These investigators fit optically thin thermal emission models to M82 and
 NGC 253 and obtained temperatures $T_X\simeq 5.7\pm 0.3$ keV (M82)
 and $6.0\pm 0.8$ keV (NGC253).
They report equivalent widths for the iron emission line at 6.7 keV
 of $170 \pm60$ eV(M82) and $180 \pm^{210}_{180}$ eV (NGC253).
It came as a surprise that there is no definitive evidence of
 iron line emission in either system, because if the X-ray emission is
 optically thin, thermal, and comes from the hot gas created by supernova
 explosions in starburst galaxies, the iron K$\alpha$ emission line $h \nu
 \simeq 6.7$ keV should have been detected.
The expected equivalent width is approximately $\sim$ 900 eV,
 if we assume that the collisional ionization balance is achieved
 and that the gas has a temperature determined by fitting the observed
 continuum spectra (Ohashi \etal 1990a).
The failure to detect the iron emission may indicate that the iron
 abundance is less than 1/5 -- 1/4 of solar abundance or that some
 non-equilibrium ionization process has invalidated some of the underlying
 assumptions.

To estimate the expected Fe abundance, we note that a type II supernova
 progenitor with an intial mass of 20 $M_\odot$ ejects $\simeq 0.1
 M_\odot$ (Thielemann, Hashimoto, and Nomoto 1990) of iron.
As adopted in our models, this amount of iron becomes mixed with 30
 $M_\odot$ of material, leading to a metalicity that is several times
 the solar value.
The amount of material that becomes mixed with the ejecta (30
 $M_\odot$) cannot be significantly larger (300--500
 $M_\odot$ would be needed to properly reduce the observed Fe
 abundance) or it would create an X-ray continuum luminosity
 in excess of that observed.

We also consider whether including non-equilibrium ionization effects can
 affect the iron abundance inferred from the observations.
Although non-equilibrium ionization models have not been developed for
 starburst galaxies, such models have been developed for supernova
 remnants, for which the physics is quite similar (Itoh 1977, 1978, and 1979;
 Shull 1982; Hamilton, Sarazin, and Chevalier 1983).
{}From hot plasma, the X-ray emission at energies $h\nu \alt$ 1keV comes
 from the line emission of He-like low-$Z$ ions such as C, N, and O.
Higher-energy X-ray emission is dominated by the continuum emission
 from Hydrogen and Helium (free-free emission).
When ionization equilibrium is not achieved (i.e., the atoms are in an
 ionization state lower than expected under equilibrium conditions),
 low-$Z$ ions emit above their equilibrium value.
In contrast, the emissivities of high-$Z$ ions, such as Fe and Ni, become
 lower than their equilibrium values.
This difference occurs because, for temperatures near $\sim 10^7$ K,
 the equilibrium ionization state of these low-$Z$ species is higher than
 for the He-like species.
The numerical result of Hamilton \etal (1983) show that the luminosity of
 Fe (and Ni) K$\alpha$ line can be less than 0.1 times that expected from
 a plasma in ionization equilibrium.
The departure from the equilibrium values depends upon the ionization
 time scale $n_0 \tau$, where $n_0$ and $\tau$ represent the ambient
 density around the supernova remnant and time elapsed since the
 explosion.
For constant values of $n_0 \tau$ (see their equation 4b and figure 6), we see
 that when $n_0 \tau=10^3 {\rm yr~cm^{-3}}$, the ratio of emission
 in the non-equilibrium stage to that in the collisional equilibrium stage,
 $F$, is approximately equal to 0.3 for a shock temperature of
 $T_s\sim 10^7$K.
In an earlier phase, where $n_0 \tau=500 {\rm yr~cm^{-3}}$, $F$
 has a value of $\alt$0.1.
Thus, it appears that if the gas has not had sufficient time to achieve
 ionization equilibrium, the 6.7 keV iron line emission may be much lower
 than value expected from ionization equilibrium, which might explain the
 small equivalent widths of iron K$\alpha$ emission lines in starburst
 galaxies.

If the average gas density in a starburst nucleus is $\sim 10^{-3}\cc$,
 the ejected matter travels a distance $\sim 1$kpc before
 achieving ionization equilibrium ($t \sim$ 1 Myr).
This situation is much different from the interstellar space in nomal
 galaxies, where all the supernova remnants except for extraordinary
 young remnants with an age of $\alt$ 1000 yr
 achieve the ionization equilibrium.
The ionization equilibrium may have been achieved in a lower halo
 or a disk  and the outflowing matter may be in the equilibrium.
However, the X-ray emission from the starburst galaxy seems to be different
 from that expected from the emission of supernova remnants in our galaxy.

\Subsubsec{4.2.b. Iron Line Emission from the Starburst Outflow}

Although the spectrum of the X-ray emission coming from the starburst
outflows
 has not been observed, here we estimate the strength of the iron line of
 the large-scale outflow.
We estimate the effective ``$n_0 \tau$'' for the starburst outflow
 by integrating the quantity
$$
\int_0^{0.95z_s} \frac{n(z)}{v_z} z^2 dz/
 \int_0^{0.95z_s} z^2 dz,
\eqno(4.2)
$$
along the $z$-axis.
The factor $\propto z^2$ means that the value should be volume averaged.
For the Sedov solution, this average value equals 4.17$n_0 \tau$.
This quantity gives us a mean measure of how close the plasma is to a
 state of equilibrium ionization.
Here, since in starburst galaxies the ejected gas seems to escape more
freely than ordinary disk galaxies, we assume that the ejected gas has a
lower ionization state than expected from its temperature using collisional
equilibrium when it leaves the galaxy.

We plot equation (4.2) as a function of time for model E in figure 8.
Since the undisturbed medium is static,
 the value $ndz/v_z$ goes to infinity in the pre-shock medium.
The gradually increasing part corresponds to the outflow material.
{}From this figure, we see that the integrated value is large
 $\sim 1.2\times 10^4\,{\rm yr~cm}^{-3}$ during the early phase ($t=$5
 Myr).
But after the outflow breaks out of the gas disk completely ($>$8 Myr),
 the swept up material is of much lower density, so the integrated value
 decreases and is not more than $\simeq$ 6000 yr cm$^{-3}$.
Since the ejected matter after the break-out can flow with slight interaction
 of ambient material, the integrated value decreases $\simeq$
 6000 yr cm$^{-3}$.
This low value seems to indicate a low degree of equilibriation.
In a late phase, the value becomes large again due to its large size and
 long age.

If we use the conversion factor from the value of equation (4.2) to $n_0 \tau$
for the
 Sedov solution, the value $\simeq 6000 {\rm yr~cm^{-1}}$
 corresponds to $\sim 1500$ yr~cm$^{-3}$ for $n_0 \tau$.
This implies that the material outflowing from a
 starburst galaxy shows modest non-equilibrium effects that could affect
 the interpretation of the iron abundance based upon the Fe K$\alpha$ line.
If we estimate the ratio of the emission in the non-equilibrium to
 the equilibrium state ($F$) based upon this value of $n_0 \tau$ as
 applied to the supernova remnant calculations ($T\sim 10^7$K)
 of Hamilton \etal (1983), we find that $F = 0.1-0.3$.
More exact simulations, i.e., 2-D hydrodynamical simulation with the
 non-equilibrium ionization, are necessary to more accurately answer the
problem, a calculation we hope to undertake in the near future.

\vspace{10mm}
One of the authors (K.T.)
 would like to thank all the members of the Astronomy Department,
 University of Michigan, Ann Arbor, for their hospitality and kindness,
 where he began this work.
The numerical calculations were carried out by supercomputers:
 {\sc Hitac} S-820/80 at the Computing Center, University of Tokyo,
 {\sc Facom} VP200E at Nobeyama Observatory, National Astronomical
Observatory,
 and {\sc Facom} VP200 at Institute of Space and Aeronautics Sciences.
This work was supported in part by the Grant-in-Aid (02854016) from
 the Ministry of Education, Science and Culture and by grant NAGW-2135
 from National Aeronautics and Space Administration.

\newpage
\Sec{Tables}
\begin{table}[h]
\centering
TABLE 1: Model Parameters

\begin{tabular}{ccccccccc}
\multicolumn{9}{c}{ } \\
\hline \hline
Model & $r_{\rm SN}$ &  $\dot{M}_{\rm SN}$ & $n_{c0}$ &
 $n_{h0}$ & $c_{sc}$ & $c_{sh}$ & spacing & number of cells\\
 & (yr$^{-1}$) & ($10^{26}{\rm g~s}^{-1}$) & ($\cc$) & ($\cc$) &
 (${\rm km~s^{-1}}$) & (${\rm km~s^{-1}}$) & (pc) & \\
\hline
A \dotfill\ & 0.1 & $1.9$ & 20 & $2\times 10^{-3}$ & 30 &
 300 & 50$\times$50 & 500$\times$250\\
&&&&&&&& 1000$\times$500\\
A1 \dotfill\ & 0.1 & $1.9$ & 20 & $2\times 10^{-3}$ & 30 &
 300 & 25$\times$25 & 500$\times$250 \\
B \dotfill\ & 0.1 & $1.9$ & 20 & $2\times 10^{-3}$ & 30 &
 $95$ & 50$\times$50 & 1000$\times$500 \\
C \dotfill\ & 0.1 & $1.9$ & 20 & $2\times 10^{-3}$ & 30 &
 $950$ & 50$\times$50 & 500$\times$250 \\
D \dotfill\ & 0.1 & $1.9$ & 20 & $2\times 10^{-4}$ & 30 &
 950 & 50$\times$50 & 500$\times$250 \\
E \dotfill\ & 0.1 & $0.3$ & 20 & $2\times 10^{-3}$ & 30 &
 $95$ & 50$\times$50 & 500$\times$250 \\
\hline
\end{tabular}
\end{table}

\newpage
\Sec{References}

\re
Chevalier, R. A., and Clegg, A. W. 1985, {\it Nature}, 317, 44

\re
Fabbiano, G. 1988, \ApJ, 330, 672

\re
Fabbiano, G., and Trinchieri, G. 1984, \ApJ, 286, 497

\re
Hamilton, A. J. S., Sarazin, C. L., and Chevalier, R. A. 1983, \ApJS, 51, 115

\re
Itoh, H. 1977, \PASJ, 29, 813

\re
------. 1978, \PASJ, 30, 489

\re
------. 1979, \PASJ, 31, 541

\re
Kronberg, P. P., Bierman, P., and Schwab, F. R. 1985, \ApJ, 291, 693

\re
Kronberg, P. P., and Sramek, R. A. 1985,  {\it Science}, 227, 28

\re
Makishima, K., and Ohashi, T. 1989, in Big Bang, Active Galactic Nuclei
 and Supernovae, ed. S. Hayakawa, and K. Sato (Universal Academy Pr.,
Tokyo),
 p.371

\re
Nakai, N. 1986, Ph.D thesis, University of Tokyo.

\re
Norman, M. L., and Winkler, K.-H. A. 1986, in Astrophysical
 Radiation Hydrodynamics, ed. K.-H. A. Winkler, and M. L. Norman
 (Reidel, Dordrecht), p.187

\re
Ohashi, T., Makishima, K., Mihara, T., Tsuru, T, Awaki, H.,
 Koyama, K., Takano, S., and Kondo, H. 1990a, in {\it Windows on Galaxies}, ed.
 G. Fabbiano, J. S. Gallagher, and A. Renzini (Reidel, Dordrecht), p.243

\re
Ohashi, T., Makishima, K., Tsuru, T., Takano, S., Koyama, K., and
 Stewart, G. C. 1990b, \ApJ, 365, 180

\re
Raymond, J. C., Cox, D. P., and Smith, B. F. 1976, \ApJ, 204, 290

\re
Shull, J. M. 1982, \ApJ, 262, 308

\re
Sakashita, S., and Hanami, H. 1986, \PASJ, 38, 879

\re
Sofue, Y., Reuter, H.-P., Krause, M., Wielebinski, R., and Nakai, N.
  1992, \ApJ, 395, 126.

\re
Telesco, C. M. 1988, \ARAAp, 26, 343

\re
Tenorio-Tagle, G., and Bohdenheimer, P. 1988, \ARAAp, 26, 145

\re
Thielemann, F.-K., Hashimoto, M., and Nomoto, K. 1990,
 in {\it Supernovae: The Tenth Santa Cruz Summer Workshop in Astronomy
 and Astrophysics}, ed. S. E. Woosley (Springer,New York), p.609

\re
Tomisaka, K. 1991, in {\it IAU Symposium 144: The Interstellar Disk-Halo
Connection
 in Galaxies}, ed. H. Bloemen, (Reidel, Dordrecht), p.407

\re
------.  1992, \PASJ, 44, 177

\re
Tomisaka, K., and Ikeuchi, S. 1988, \ApJ, 330, 695 (Paper I)

\re
Tsuru, T., Ohashi, T., Makishima, K., Mihara, T., and Kondo, H. 1990,
 \PASJ, 42, L75

\re
van Leer, B. 1977, {\it J. Comp. Phys.}, 23, 276

\re
Watson, M. G., Stanger, V., and Griffiths, R. E. 1984, \ApJ, 286, 144

\re
Weaver, R., McCray, R., Castor, J., Shapiro, P., and Moore, R. 1977, \ApJ,
 218, 377 (errata 220, 742).

\newpage
\Sec{Figure Captions}

\re
{\bf Fig.1:} (a) The initial state of the
 density (left panel) and pressure (right panel) distributions.
The disk component occupies the region below $ z_d$, while an extensive
 halo of nearly contant density lies above it.
(b) The initial distribution of pressure (dash-dotted line) and
 density (dashed line) are plotted along the $z$-axis (left panel)
 and the $r$-axis (right panel), where the standard pressure, $p_*$,
 is taken $10^4~{\rm K~cm^{-3}}$.

\re
{\bf Fig.2:} The structure of the starburst outflow for model A at four
 times:  $t=5$Myr (a), $t=10$Myr (b), $t=20$Myr (c), and $t=50$Myr (d).
The right panel shows pressure contours and velocity vectors while
 the left panel shows density contours.
Before $t=$5Myr, the bubble is already focused by
 the effect of the initial density gradient,
 and after $t=$8Myr, the bubble expands in a uniform-density halo,
 producing shocks in the process.

\re
{\bf Fig.3:} Cross-sectional views of the starburst outflow
 along the $z$-axis (left panel) and the $r$-axis (right panel),
 showing velocity $\log |v_z/v_*|$ (solid line),
 temperature $\log T$(K) (dash-dotted line),
 density $\log n{\rm (cm^{-3})}$ (dashed line),
 and Mach number $M_z=v_z/(\gamma p/\rho)$.
Here, $v_*=11.65 {\rm km~s^{-1}}$, is the isothermal sound speed for $10^4$K
gas.
Figures (a) - (d) correspond to (a) - (d) of figure 2, respectively.
The heights of the cross-section views (right panel) are at
 $z=0$kpc (a), 5kpc (b, c), and 10kpc(d).

\re
{\bf Fig. 4}: The expansion law of the outflow, showing the time variation
 of the outermost shock front traveling on the $z$-axis.
This clearly shows that the expansion is divided into three phases.
First, a hot bubble is created in the high density disk, and
 the shock decelerates.
Second, when the size of the bubble surpasses the size of the disk,
 it elongates in the $z$-direction and the shock acclerates into the
 stratified density distribution.
Finally, the shock begins to decelerate when it encounters the
 uniform-density halo.

\re
{\bf Fig.5}: (a) The structure of the starburst outflow for model B at
 $t=50$Myr, where the right panel shows pressure contours and
 velocity vectors while the left panel shows density contours.
This model has the same density distribution as model A but has a
 pressure and a temperature 10 times smaller.
Comparing this with figure 2d, the expansion in the $z$-direction is
 almost identical, but the expansion in the $r$-direction differs
 by $\simeq$ 40\%.
(b)  Cross-sectional views of the starburst outflow
 along the $z$-axis (left panel) and the line $z=10$kpc (right panel),
 showing velocity $\log |v_z/v_*|$ (solid line),
 temperature $\log T$(K) (dash-dotted line),
 density $\log n{\rm (cm^{-3})}$ (dashed line),
 and Mach number $M_z=v_z/(\gamma p/\rho)$.

\re
{\bf Fig.6}: The time variations of X-ray luminosity from
 the outflowing hot gas (model B) are given in the energy bands
 0.155--24.8 keV (solid line), 0.284--1.56 keV (dotted line),
 and 1.56--8.265 keV (dashed line).
The second and third band correspond, respectively, to those of the
 Einstein Observatory HRI and the Ginga LAC.

\re
{\bf Fig.7}: The same as figure 6 but for model E.
Here, the mass loss rate from the supernovae plus evolving stars is 1/6
 that of model B; all other parameters are the same as in model B.

\re
{\bf Fig.8}: The distribution of $\int_0^z n dt$ along the $z$-axis for
 model E.
Since the initial halo material is static, the value of $\int_0^z n dt$
 is artificially large at early times.
However, after 8 Myr, the value of $\int_0^z n dt$ is never more than
 $\simeq$ 6000 yr cm$^{-3}$, so that the emitting gas is in a modestly
 non--equilibrium state.
This may help to explain the low observed value for the equivalent widths
 of the iron K$\alpha$ emission lines.

\newpage}
\end{document}